# A note on Gibbs paradox

P. Radhakrishnamurty

282, Duo Marvel Layout, Anathapuragate, Yelahanka,

Bangalore 560064, India. email: padyala1941@yahoo.com

Abstract

We show in this note that Gibbs paradox arises not due to application of thermodynamic principles, whether classical or statistical or even quantum mechanical, but due to incorrect application of mathematics to the process of mixing of ideal gases.

-----------------------------------------------------------------------------------------------------------------------

## 1. Introduction

'It has always been believed that Gibbs paradox embodied profound thought', Schrodinger [1].

It is perhaps such a belief that underlies the continuing discussion of Gibbs paradox over the last hundred years or more. Literature available on Gibbs paradox is vast [1-15]. Lin [16] and Cheng [12] give extensive list of some of the important references on the issue, besides their own contributions.

Briefly stated, Gibbs paradox arises in the context of application of the statistical mechanical analysis to the process of mixing of two ideal gas samples. The discordant values of entropy change associated with the process, in the two cases: 1. when the two gas samples contain the same species and, 2. when they contain two different species constitutes the gist of Gibbs paradox.

A perusal of the history of the paradox shows us that: Boltzmann, through the application of statistical mechanics to the motion of particles (molecules) and the distribution of energy among the particles contained in a system, developed the equation: $S = \log W + S(0)$, connecting entropy $S$ of the system to the probability $W$ of a microscopic state corresponding to a given macroscopic equilibrium state of the system, $S(0)$ being the entropy at 0K. Boltzmann did not assign any value to $S(0)$. Planck modified the equation by introducing the constant, $k$, which we now call the Boltzmann constant, and setting $S(0)$ to zero, and wrote: $S = k \log (W)$ [17]. Gibbs found that this equation when applied to the calculation of the entropy of an ideal gas led to results that showed the nature of Boltzmann's statistical entropy was not in harmony with the nature of classical thermodynamic Clausius entropy. Gibbs offered a resolution of the issue (that we now call Gibbs paradox) by proposing a correction to the method of counting the number of microstates that correspond to a given macro state, to obtain the value of $W$ of a system of two ideal gases, when the particles in the two gases are distinguishable or are indistinguishable. This correction led to a debate of the issue ever since.

## 2. Entropy of ideal gas mixing process

The change in entropy, $\Delta S$, of the process of mixing two ideal gases, A and B, that are at the same temperature, T, is given by the equation (1) found in any standard book on the subject [1- 5].

$$\Delta S = -N_A k \ln X_A - N_B k \ln X_B = -\sum_j N_j k \ln X_j = -Nk \sum_j X_j \ln X_j, \quad j = A, B \tag{1}$$

$N_j$ represents the number of molecules and $X_j$ represents the mole fractions of the species j in the mixture. Since X is a proper fraction, $\Delta S > 0$. Both classical thermodynamics and statistical mechanics give equation (1) [2, 3], when applied to the mixing process.

In the special case, when $N_A = N_B = N$, we get, $\Delta S = 2Nk \ln 2 > 0$

Equation (1) applies to both the cases: (i) when the pure components and the mixture have the same values of pressure P and temperature T, that is when, $\sum V_j = V_t$, $\sum N_j = N_t$, $P_j = P$, $T_j = T$ and, (ii) when the pure components and the mixture have the same values of volume V and temperature T, that is when, $\sum P_j = P_t$, $\sum N_j = N_t$, $V_j = V$, $T_j = T$, depicted in Fig. 1. In case (i), Amagat's law[18], eq (2) below applies and in case (ii) Dalton's law [18], eq (3) below, applies to the ideal gas mixtures. $V_j$, $P_j$ are respectively, the volumes and pressures of the species j before mixing. $P_j$ and $V_j$ are usually known as partial pressure and partial volume of j in the mixture. $N_t$ is the total number of molecules in the system.

$$\text{Amagat's law: } \frac{N_j}{\sum_j N_j} = \frac{N_j}{N_t} = \frac{V_j}{\sum_j V_j} = \frac{V_j}{V_t} = X_j, \quad j = A, B \tag{2}$$

$$\text{Dalton's law: } \frac{N_j}{\sum_j N_j} = \frac{N_j}{N_t} = \frac{P_j}{\sum_j P_j} = \frac{P_j}{P_t} = X_j, \quad j = A, B \tag{3}$$

When the system undergoes a process that takes the system from an initial equilibrium state i to a final equilibrium state f, the entropy of the system in the initial state, $S^i$, the entropy in the final state, $S^f$, and $\Delta S$ are given respectively, by the equations (4), (5) and (6) below.

$$S^i = -N_A k \ln X_A^i - N_B k \ln X_B^i \tag{4}$$

$$S^f = -N_A k \ln X_A^f - N_B k \ln X_B^f \tag{5}$$

$$\Delta S = (S^f - S^i) = -N_A k \ln\left[\frac{X_A^f}{X_A^i}\right] - N_B k \ln\left[\frac{X_B^f}{X_B^i}\right] = -Nk \sum_j X_j \ln\left[\frac{X_j^f}{X_j^i}\right] \tag{6}$$

## 3. Initial state of the system

It is important to note that in the process of mixing under consideration here, the initial state of the system corresponds to the pure components A and B, separated from each other by a partition (see Fig. 1).

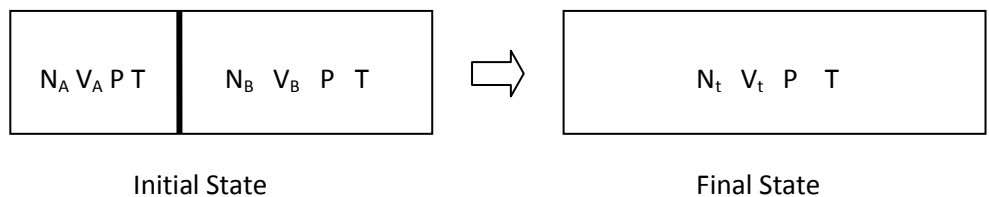

Initial State          Final State

Fig. 1a. A partition separates the ideal gases A and B in the initial state. The final state is the mixture of the two gases with no partition. Amagat's law applies to the mixture. The system of the two gases is an isolated system. $(V_A+V_B)=V_t$.

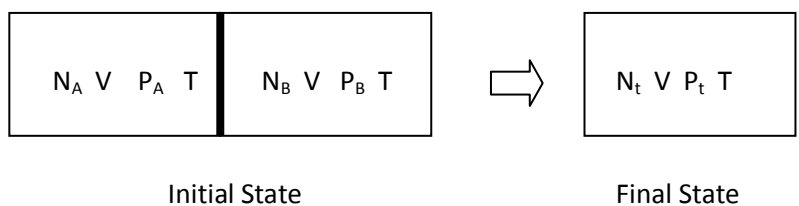

Initial State          Final State

Fig. 1b. A partition separates the ideal gases A and B in the initial state. The final state is the mixture of the two gases. Dalton's law applies to the mixture. The system of the two gases is a closed system. $(P_A+P_B)=P_t$.

## 4. Importance of constraints governing the mixing process

Therefore, for this special (not an arbitrary) initial state case, we have,

$$X_A^i = X_B^i = X_j^i = 1 \tag{7}$$

However, in the final equilibrium state, we have,

$$(X_A^f + X_B^f) = 1 \quad or \quad X_A^f = (1 - X_B^f) < 1 \quad or \quad X_j^f < 1 \tag{8}$$

We note that eqs (7) and (8) are mutually exclusive and do not hold good together (simultaneously).

To show the constraints imposed by eqs (7) and (8) on eqs (4) and (5) explicitly, we write eqs (4) and (5) as eqs (9) and (10) respectively, indicating the values $X_i s$ can take in the initial and final states.

$$S^i = -N_A k \ln X_A^i - N_B k \ln X_B^i, \quad (X_j^i = 1) \tag{9}$$

$$S^f = -N_A k \ln X_A^f - N_B k \ln X_B^f, \quad (X_j^f < 1) \tag{10}$$

We see that eqs (9) and (10) make incompatible demands on $X_j$. As such, combination of eqs (4) and (5) to get $\Delta S$, in this special case, leads to mathematical inconsistencies that lead to Gibbs paradox.

Therefore, mathematical consistency demands that eqs (4) and (5) can be combined, if and only if, both of them are constrained by either one of the two eqs (7) or (8), lest a paradox arise. We can conveniently discuss the different possibilities that arise in the two cases, by dividing them into two broad categories.

I. $X_j^i = X_j^f = 1$

In this case, combination of eqs (4) and (5) is constrained by eq (7). This constraint demands that the initial and final states of the system correspond to a pure substance only, no mixture. Therefore, the species on both sides of the partition are indistinguishable (A= B). Hence the subscripts and superscripts play no role and can be dropped to give X = 1. The question of mixing does not arise at all. Therefore, $\Delta S = 0$, in this case. This result shows that removal of the partition is a reversible process. We can reintroduce the partition with $\Delta S = 0$, for the process of reintroducing the partition. T and P of the system are same before and after removal of the partition.

II. $X_j^i = X_j^f < 1$

In this case, combination of eqs (4) and (5) is constrained by eq (8) and the initial and final states correspond to mixture only (no pure components in either of the states). T and P of the system are the same before and after mixing. We get several possibilities again here.

II(a). $X_{jl}$ = constant < 1. The second subscript is used to denote the number of the mixture (l = 1, 2, ……). There are three mixtures – two mixtures before the partition is removed and one mixture (the mixture of mixtures) after the partition is removed. In this case, all mixtures have the same composition. Thus removal of partition leads to no change of mole fraction of any species leading to $\Delta S = 0$, in this case also.

II(b). $1 > X_{jl} \neq X_{jm} < 1$.

This is the most important case. In this case no two mixtures have the same composition. They all have the same temperature and pressure but different compositions. When two such mixtures are mixed by removing the partition separating them, the resulting mixture will have a composition different from that of either of the initial mixtures; but the change of the composition entails no change in entropy of the system, lest it violate Amagat's law; consequently violate the ideal gas law. The system can be brought back to the initial state by using membranes that allows only specific species to pass through them [6].

Similar arguments apply to the case when the temperature and volume (instead of pressure) of the three mixtures is the same, but compositions are different (note, the system is a closed system in this case. The process connecting the initial and final states is not just a mixing step but includes other reversible steps. The entropy change of the system in the reversible steps can be accounted for). The mixture of mixtures will have a composition different from that of either of the initial mixtures; but again, the change of composition entails no change in entropy of the system, lest it violate Dalton's law; consequently violate the ideal gas law.

This is an excellent demonstration of the fact that a mixture at a given temperature and pressure (or volume) has no unique equilibrium composition. (Recall, we teach children in school that one of the main differences that distinguishes a mixture from a compound is that while a compound has a definite composition, a mixture has none). Mixtures of ideal gases having the same temperature, pressure and a given value of total number of molecules can have a large number of different compositions. All such mixtures are governed by Amagat's law and exist in mutual equilibrium. None of the component mixtures is more stable than any other component mixture or is less stable than their mixture! They all exist in mutual equilibrium in spite of the fact that they have different compositions; there exists no equilibrium composition they can seek, different from the one they already are in! Similar arguments apply to mixtures obeying Dalton's law. Composition is not a criterion of equilibrium between ideal gas mixtures.

These ideal gas mixtures have the same value of total entropy before mixing as the entropy they have after mixing. Therefore, the change in entropy of a system of ideal gas mixtures due to the process of mixing is zero. This is the essence of equation (6). In fact, it says more than this. It says, the entropy change of the mixing process can be calculated as if molecules of the same type are mixing (that is, type A with type A, type B with type B and so on, so that no entropy change occurs due to the mixing process! It is a consequence of the fact that each term in equation (6) is zero.

Thus the above analysis shows that no paradox arises if eqs (4) and (5) are combined when, and only when, both of them are constrained either by eq (7) or by eq (8).

Therefore, it is evident that whether A and B are the same species or different species matters little for the change in entropy of the process of mixing. In fact, it doesn't even matter if A and B are pure substances or mixtures! This is in contrast to the prediction of the statistical mechanical result that the value of $\Delta S$ depends on whether A and B are same or not (distinguishable or indistinguishable).

## 5. Discussion

The foregoing analysis demonstrates that Gibbs paradox arises not due to application of thermodynamic considerations - whether classical or statistical or even quantum mechanical – but due to incorrect application of mathematics to the process of mixing of ideal gases. The paradox is not connected with the nature of properties of extensivity and additivity of entropy. Efforts to apply corrections to the Boltzmann counting process by way of introducing the factor N! to arrive at the correct results, cannot remove the paradox. The debate on this issue of inclusion of N! is irrelevant to the paradox. The arguments, that there is no real paradox in the Gibbs paradox, as evidenced by the results of application of quantum mechanics to the mixing process, do not hold either.

## 6. Comments

It may not be out of place to make the following comments here. The beauty of symmetry in equations (4) and (5), which remain unaltered with interchange of the superscripts i, f, shows the irrelevance of the role of considerations of time for the process – a Sterling test for thermodynamic processes! Again, the symmetry with respect to the interchange of the subscripts A and B, which leaves the equations unaltered shows that distinguishability or otherwise of A and B is irrelevant for the analysis of Gibbs paradox. In fact it highlights the irony in the very concept of *plurality of ideal gases*. The designation of the species by A and B is against the spirit of the concept of an ideal gas. There exists but one ideal gas! The ideal gas law: PV =NRT, contains N, the number of moles, but not the molecular weight – a characteristic property of a gas; thereby ridding itself from the chemical nature of the gas. When there exist no more than one ideal gas, the mole-fractions, $X_j$, lose their relevance.

## 7. Acknowledgement

I thank Director, Raman Research Institute, Bangalore, India, for extending me library facilities. I thank Arun Mozhi Selvan who constantly supports and encourages my research work.